\documentclass[conference]{IEEEtran}
\IEEEoverridecommandlockouts
\usepackage{cite}
\usepackage{amsmath,amssymb,amsfonts}
\usepackage{algorithmic}
\usepackage{graphicx}
\usepackage{textcomp}
\usepackage{xcolor}
\def\BibTeX{{\rm B\kern-.05em{\sc i\kern-.025em b}\kern-.08em
    T\kern-.1667em\lower.7ex\hbox{E}\kern-.125emX}}

\usepackage{booktabs}
\usepackage{listings}
\usepackage{amsmath}
\usepackage{url}
\usepackage{float}
\usepackage{hyperref}

\lstdefinestyle{python}{
    language=Python,
    basicstyle=\ttfamily\small,
    keywordstyle=\color{blue!70!black}\bfseries,
    commentstyle=\color{gray},
    stringstyle=\color{orange!70!black},
    showstringspaces=false,
    breaklines=true,
    breakatwhitespace=false,
    frame=single,
    numbers=none,
    columns=fullflexible,
    tabsize=4,
    backgroundcolor=\color{gray!5}
}
\begin{document}

\title{Accelerated Genetic Programming Hyper-Heuristics for Simulation-Based Scheduling via Agentic AI}


\author{\IEEEauthorblockN{Heyang Thomas Li}
\IEEEauthorblockA{\textit{Research Engineering Team}\\
\textit{REANNZ}\\
New Zealand \\
thomas.li@reannz.co.nz \\
heyang.li@anu.edu.au}
\and
\IEEEauthorblockN{Alexander Pletzer}
\IEEEauthorblockA{
\textit{Research Engineering Team}\\
\textit{REANNZ}\\
New Zealand \\
alexander.pletzer@reannz.co.nz}
\and
\IEEEauthorblockN{Yuan Tian}
\IEEEauthorblockA{
\textit{Centre for Data Science and Artificial Intelligence} \& \\ \textit{School of Engineering and Computer Science}\\
\textit{Victoria University of Wellington}\\
New Zealand \\
yuan.tian@vuw.ac.nz}
\and
\IEEEauthorblockN{Yi Mei}
\IEEEauthorblockA{
\textit{Centre for Data Science and Artificial Intelligence} \& \\ \textit{School of Engineering and Computer Science}\\
\textit{Victoria University of Wellington}\\
New Zealand \\
yi.mei@vuw.ac.nz}
\and
\IEEEauthorblockN{Mengjie Zhang}
\IEEEauthorblockA{
\textit{Centre for Data Science and Artificial Intelligence} \& \\ \textit{School of Engineering and Computer Science}\\
\textit{Victoria University of Wellington}\\
New Zealand \\
mengjie.zhang@vuw.ac.nz}
}

\markboth{}{}

\maketitle

\begin{abstract}
Python is widely used in scientific research because it enables rapid development and provides rich ecosystems for data analysis, artificial intelligence (AI), and machine learning. However, customized research code can become prohibitively slow as experiments scale. This challenge is particularly acute in discrete-event project-scheduling simulations, where sequential state updates, nested loops, conditional evaluations, and object-oriented structures limit the benefits of compiled numerical and GPU-accelerated libraries. Addressing these bottlenecks typically requires iterative profiling, refactoring, testing, and validation, yet researchers may lack the time or specialized software-engineering expertise for low-level optimization. This paper presents a systematic refactoring approach using Claude agentic AI on real-world project-scheduling workloads in a high-performance computing (HPC) environment. Guided by representative benchmarks and correctness checks, the agent identifies bottlenecks, implements targeted optimizations, and evaluates their effects, while the researcher retains final control. Testing runtime reduced from 1,298 seconds to under 200 seconds without changing outputs, saving four million core-hours (NZ\$320,000) annually.

\end{abstract}


\section{Introduction}

As the Python ecosystems for data analysis, artificial intelligence (AI), and machine learning (ML) have continued to expand and mature, Python has become a preferred programming language for many researchers because of its accessibility, rapid development cycle, and extensive library support. However, its interpreted and dynamically typed execution model can impose substantial performance overhead. Optimized libraries such as NumPy \cite{harris2020array} and PyTorch \cite{paszke2017automatic} mitigate this limitation by executing computational kernels in compiled code or on GPUs. Such acceleration is less straightforward for applications with complex control flow, including discrete-event simulations, where execution involves numerous loops, conditional evaluations, state updates, and sequentially dependent events that cannot easily be expressed as vectorized array or tensor operations.
This limitation is particularly consequential when AI algorithms are used to automate the design of heuristic rules for dynamic scheduling. During training, the algorithm generates and evaluates a large number of candidate rules, each of which must be assessed through repeated simulation. Consequently, simulation often becomes the dominant computational bottleneck. Access to high-performance computing infrastructure can increase experimental throughput, but it does not eliminate the underlying technical debt: inefficient simulation code still consumes excessive CPU time and energy, limits the number and scale of experiments, and increases the cost of scientific discovery.

This performance engineering project was initiated to address severe resource exhaustion on a compute cluster managed by Research and Education Advanced Network New Zealand (REANNZ). According to REANNZ systems support team:

\begin{quote}
\itshape
``The project had been renewed for 1,054,000 compute units for the year from 2026-09-01. However, over the last year it used \textasciitilde5,173,000 compute units.''
\end{quote} \ \\
Unchecked, the project consumed roughly five times the number of compute units (a compute unit roughly matches a core-hour). This paper details the performance engineering strategies used to bring the project back into acceptable computational bounds while improving research throughput and reducing operational cost.

This situation reflects a broader tension in research software. Research emphasizes novel concepts, mechanisms, and scientific results, so implementation efficiency often becomes secondary once a proof of concept produces valid outputs. Performance engineering is also iterative and time-consuming, requiring representative benchmarks, profiling, repeated modifications, and correctness and runtime checks. Domain researchers may also have limited formal software-engineering training \cite{wilsonGoodEnoughPractices2017}.

Agentic coding assistants offer a way to lower this barrier. Claude Code \cite{anthropic_claude_2026} and CodeX \cite{openaiCodex2025} can inspect repositories, edit files, and run profilers and tests. Their emergence accompanies ``vibe coding,'' or developing software through natural-language interaction with a code-generating model \cite{sarkarVibeCoding2025}. Beyond generating code, an agent can explore alternatives, automate repeated benchmarks, compare measured effects, and discuss trade-offs with the researcher.

Agentic AI nevertheless cannot reliably optimize scientific software from a single wish-like prompt. A change may alter numerical results, move the bottleneck, or improve a micro-benchmark without reducing end-to-end runtime. Our contribution is therefore a practical human--agent workflow, demonstrated on a Genetic Programming Hyper-Heuristic (GPHH) code for simulation-based scheduling \footnote{The source code is available at \href{https://github.com/TianYuanSX/GP4MRCPSP}{github.com/TianYuanSX/GP4MRCPSP}}. A trusted benchmark establishes runtime and solution checksums; the agent profiles the program, proposes scoped changes, and tests their effects; and the researcher reviews each implementation and independently accepts or rejects it. This disciplined loop turns unconstrained code generation into evidence-based performance engineering and delivers substantial efficiency and economic gains without changing the scheduling algorithm or its outputs.

The remainder of this paper covers the scheduling and simulation background (Sec.~\ref{sec:background}), the DMRCPSP case study and GPHH decision process (Sec.~\ref{sec:DMRCPSP}), and the human--agent workflow and optimization strategies (Sec.~\ref{sec:optimization}), followed by their measured impact and conclusions.

\section{Background and Related Work} \label{sec:background}

\subsection{Societal and Economic Impact in Planning and Scheduling}

Planning and scheduling allocate scarce resources over time, with consequences beyond completion time. In emergency medical services, ambulance dispatch and relocation affect response times \cite{schmidDynamicAmbulance2012}. Genetic programming hyper-heuristics have recently produced interpretable rules for vehicle-subset selection and interdependent dispatch decisions \cite{maclachlanAmbulanceSubset2022,maclachlanEmergencyDispatch2023}. A cohort study of out-of-hospital cardiac arrests predicted that reducing the response-time target from 14 to 8 minutes would increase survival from 6\% to 8\%, while a five-minute target would increase survival to 10--11\% \cite{pellAmbulanceResponse2001}. Health-care planning also coordinates operating rooms, beds, diagnostic services, and staff, affecting access and waiting times \cite{hulshofHealthcarePlanning2012}.

The economic and environmental consequences are similarly substantial. A dynamic job-shop study reported makespan reductions of 2--17\% and machine-utilization improvements of 2--21\% after adapting dispatching rules to changing conditions \cite{zhaoRealtimeJobShop2022}; power-system load scheduling can improve both market outcomes and energy use \cite{luSmartGridScheduling2018}. This impact extends from recurring operations to projects, where scheduling coordinates interdependent activities and resources. A construction case study, for example, reported that resource-aware scheduling improved duration by 6.25\% and profit by 8\% \cite{christodoulouResourceConstrained2012}. Such settings motivate the use of a resource-constrained project scheduling problem (RCPSP) approach, a standard abstraction for resource-constrained project planning.

\subsection{RCPSP Complexity and Evolutionary Computation}

The resource-constrained project scheduling problem (RCPSP) comprises precedence-related activities with given durations and demands for limited renewable resources. It assigns start times that respect precedence and resource capacities, usually to minimize project makespan \cite{hartmannUpdatedSurveyVariants2022}. RCPSP-based models support planning in manufacturing, supply chains, software projects, and aircraft maintenance \cite{rahmanEnergyefficientProjectScheduling2022,asadujjamanSupplyChainIntegrated2024,kurtResourceConstrainedMultiproject2018,chenResourceconstrainedProjectScheduling2024}.

Even the classical RCPSP has non-polynomial (NP) complexity \cite{blazewiczSchedulingComplexity1983}. Its difficulty arises because precedence and resource constraints couple the start-time decisions: starting one activity can delay several others that require the same resources. The number of candidate activity sequences grows combinatorially with project size, while each sequence must also be checked for temporal and resource feasibility. Exact methods can therefore become impractical as the number of activities and constraints increases. Even a recent compact exact formulation for robust MRCPSP, although markedly faster than prior approaches, was tested only up to 20 activities; it solved on average 88.1\% of those instances to optimality within two hours, falling to 86.3\% at the highest uncertainty level \cite{bold2022faster}.

Evolutionary computation (EC) explores these large spaces under a finite computational budget. Competitive and hybrid genetic algorithms have produced high-quality RCPSP schedules on benchmarks \cite{hartmannCompetitiveGA1998,vallsHybridGA2008}. Genetic programming (GP) is instead used as a hyper-heuristic: it evolves reusable priority rules that construct schedules from the current state and has been effective in complex, dynamic scheduling \cite{nguyenGeneticProgrammingProduction2017,zhangSurveyGeneticProgramming2023}. Their rules are inexpensive to apply online, but training remains costly because thousands of candidates must be evaluated over many simulated instances and stochastic realizations.

\subsection{Simulation-Based Evaluation and Acceleration Challenges}

For realistic dynamic scheduling, heuristic-rule quality is often estimated using discrete-event simulation (DES) \cite{amaranSimulationOptimization2016}. Python is attractive for developing such models because its concise syntax and flexible object model support rapid prototyping and iterative refinement. Its broad ecosystem also allows data preparation, statistical analysis, graph processing, visualization, and learned heuristic rules to be implemented in one environment using packages such as pandas\cite{reback2020pandas}, NumPy \cite{harris2020array}, NetworkX \cite{hagberg2008}, and PyTorch \cite{paszke2017automatic}.

The trade-off is execution speed. CPython incurs overhead from bytecode dispatch, name and attribute lookup, dynamic type handling, reference counting, and garbage collection \cite{zhangPythonInterpretation2022}. These costs become substantial in fine-grained loops that repeatedly create objects or invoke small functions. High-level packages can introduce further overhead because their general-purpose interfaces support flexible data structures, validation, conversion, iterators, and rich abstractions. Although numerical libraries execute large kernels efficiently in compiled code, repeated small library calls, temporary arrays, and transitions between Python and native code may prevent that cost from being amortized. Sequential DES magnifies these effects by executing the event and decision logic many times.

Several approaches can accelerate Python simulation, but each has limitations. Independent replications and heuristic-rule evaluations can run in parallel, but scaling this approach requires many CPU cores and duplicates simulation state in memory. Parallel DES can distribute a single simulation, but must preserve causality through conservative or optimistic synchronization \cite{fujimotoParallelSimulation2016}. GPUs provide greater parallel throughput, yet fine-grained DES maps poorly to them because events update the state required by later decisions and cause asynchronous advancement, synchronization, irregular memory access, and branch divergence \cite{tangGPUDiscreteEvent2013}. A third approach is to compile selected performance hots pots to native code with tools such as Numba \cite{lamNumbaLLVMbasedPython2015} or Cython. This removes interpreter overhead without rewriting the entire simulator. For Numba in particular, efficient compilation requires well-defined types and supported control flow; dynamic containers, arbitrary Python objects, and object-oriented features have limited support and often must be converted into primitive values and contiguous arrays \cite{lamNumbaLLVMbasedPython2015}.

\section{Case Study Domain: GPHH for DMRCPSP} \label{sec:DMRCPSP}

\subsection{Dynamic Multi-Mode Resource-Constrained Project Scheduling Problem (DMRCPSP)}

\begin{figure}[t]
\centering
\includegraphics[trim={0 0 0 2.5cm}, clip, width=\columnwidth]{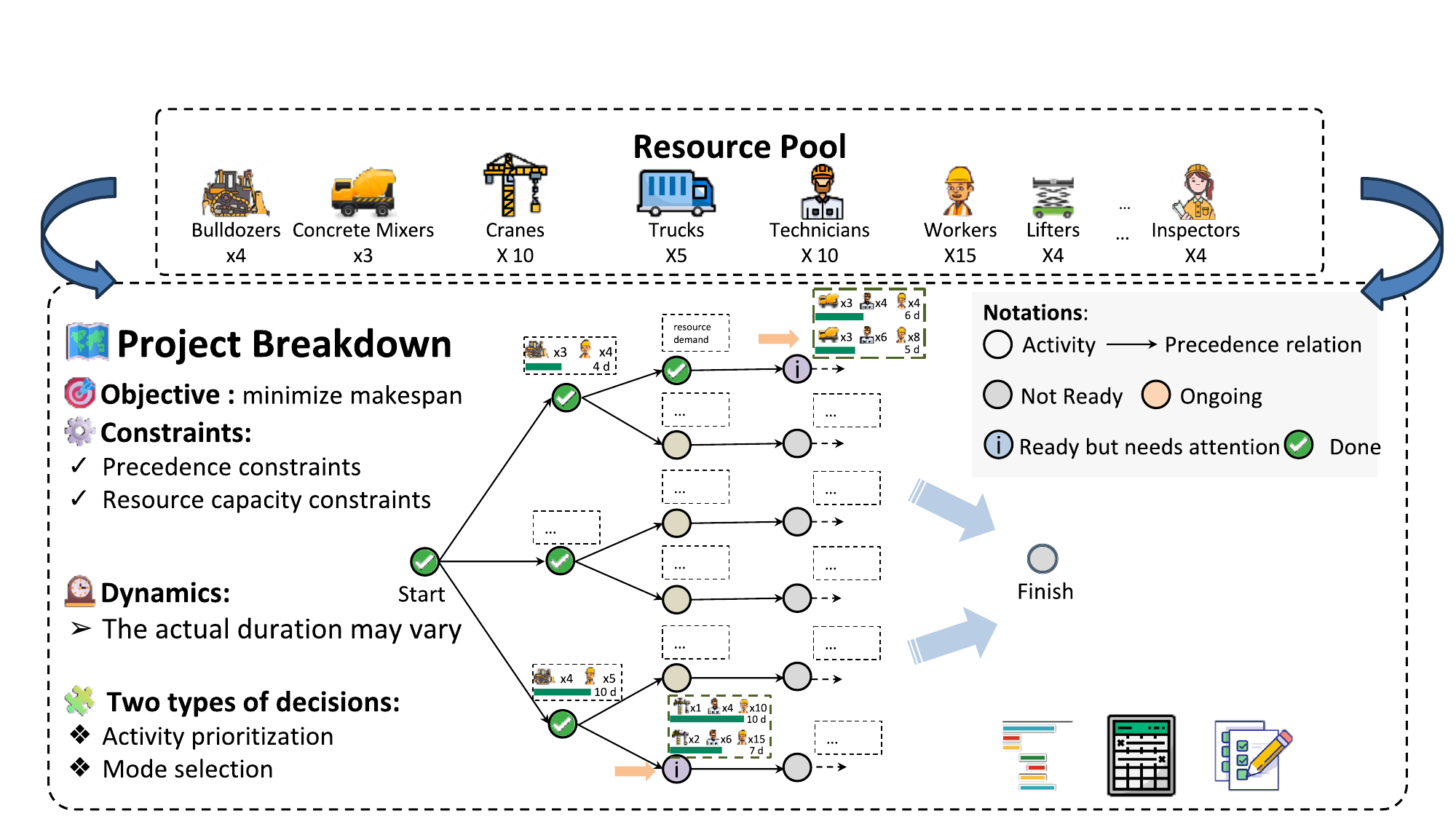}
\caption{Overview of the Dynamic Multi-mode Resource-constrained Project Scheduling Problem (DMRCPSP): a capacity-limited resource pool, a project breakdown structure with precedence and resource constraints, and stochastic activity durations.}
\label{fig:dmrcpsp}
\end{figure}

\begin{figure*}[ht]
\centering
\includegraphics[width=0.94\textwidth]{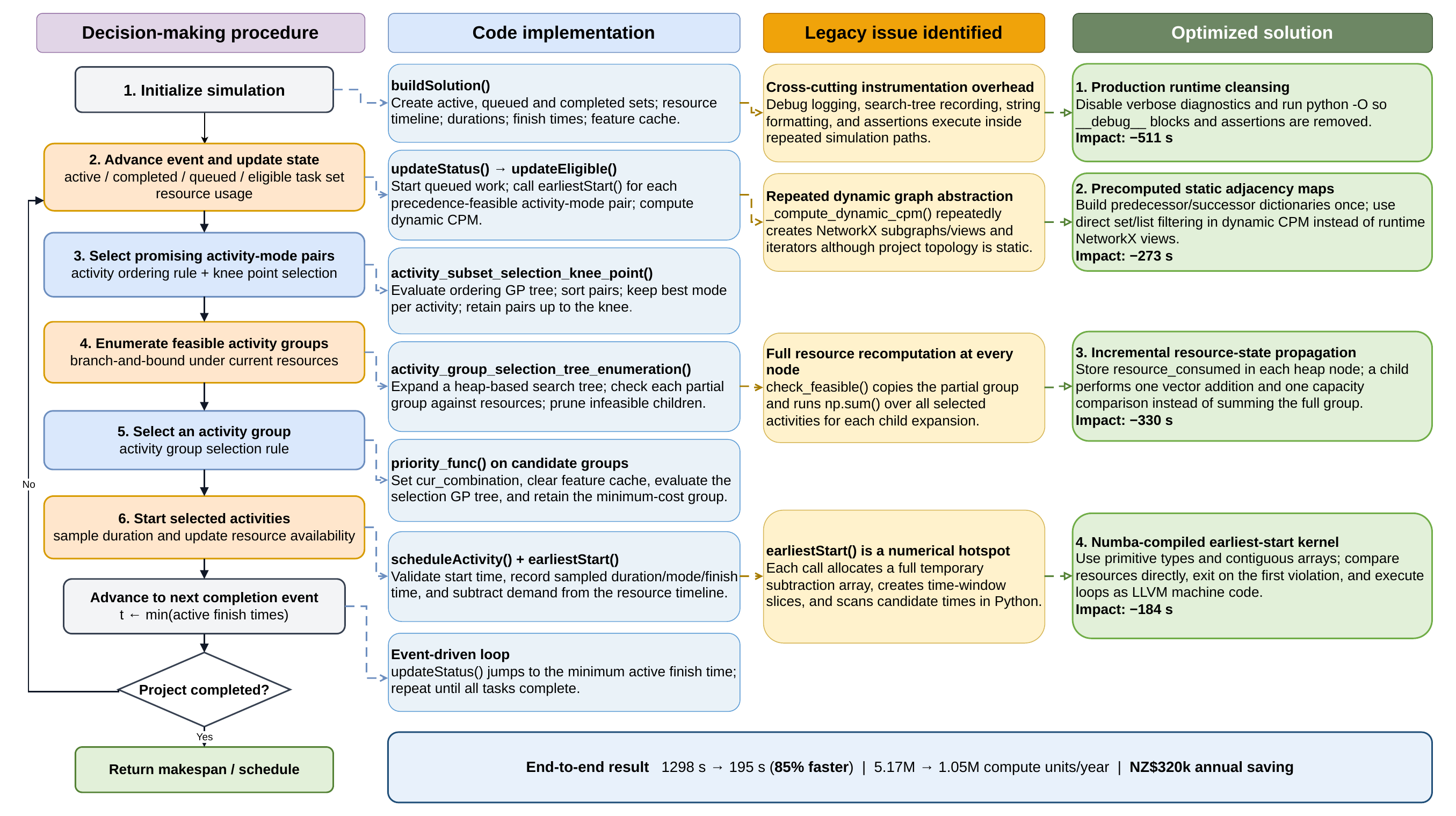}
\caption{DMRCPSP simulation workflow and summary of the Agentic-AI-assisted code optimization. The decision-making procedure is mapped to its Python implementation, the performance issues identified in the legacy code, and the corresponding optimized solutions. Solid arrows show simulation control flow; dashed arrows link related implementation, issue, and solution components.}
\label{fig:case-study-optimization}
\end{figure*}

\begin{figure*}[htbp]
\centering
\includegraphics[width=0.88\textwidth]{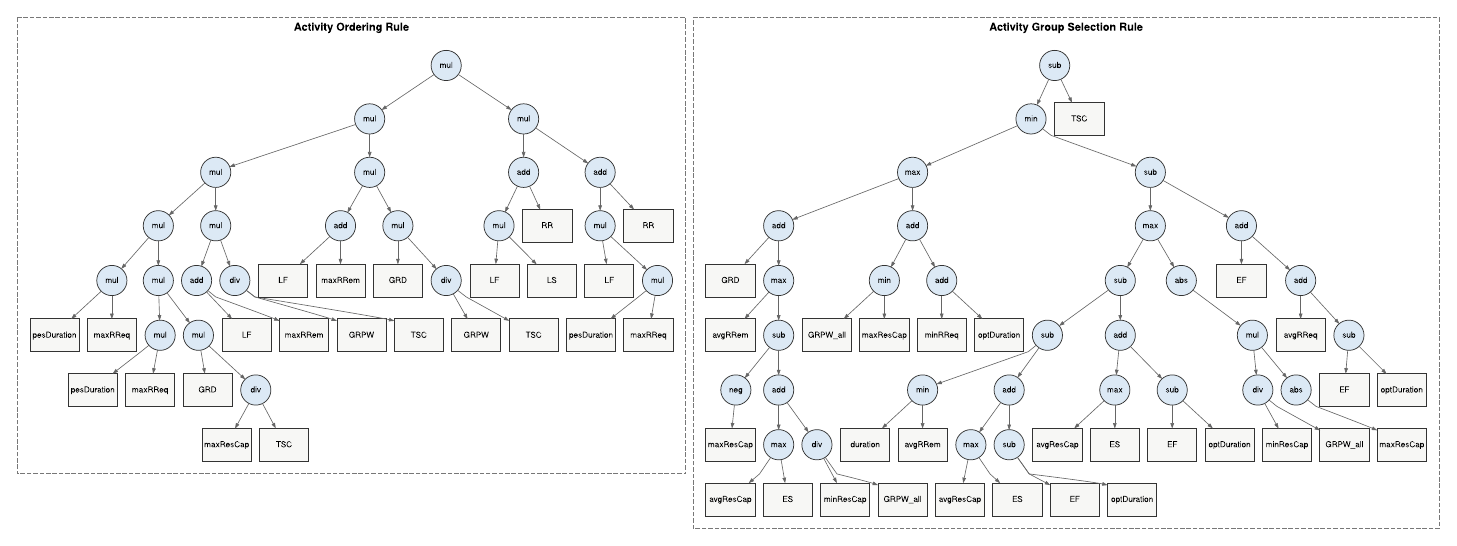}
\caption{Example of an evolved GP individual containing an activity--mode ordering rule (top) and an activity-group selection rule (bottom), each represented as a mathematical expression tree.}
\label{fig:evolved_rule}
\end{figure*}

This case study focuses on the dynamic multi-mode resource-constrained project scheduling problem (DMRCPSP), illustrated in Fig.~\ref{fig:dmrcpsp}. It extends the RCPSP in two ways. First, each activity can be executed in one of several modes with different expected durations and resource demands \cite{peteghemExperimentalInvestigationMetaheuristics2014}. Second, an activity's actual duration is uncertain and is revealed only during execution. Scheduling decisions must therefore be made online: as the project state changes, heuristic rules select feasible activity--mode combinations from the currently eligible set. The objective is to minimize expected makespan while satisfying precedence and resource constraints.

\subsection{Decision-Making Workflow and Heuristic Rules for DMRCPSP}
The DMRCPSP decision-making workflow \cite{tianScalableKneePoint2026} is shown in Fig.~\ref{fig:case-study-optimization} . This figure also connects the event-driven simulation workflow to its Python implementation and summarizes the performance issues and corresponding solutions identified through the Agentic-AI-assisted optimization process. The simulation first initializes the activity states, stochastic durations, and resource-availability timeline. At each decision point, completed activities are recorded, their resources are released, and the eligible set is updated to contain activity--mode pairs whose precedence and resource requirements permit an immediate start.

Schedule construction then makes two types of heuristic decisions, illustrated by the evolved-rule example in Fig.~\ref{fig:evolved_rule}:
\begin{enumerate}
    \item \textbf{Activity--mode ordering}: An ordering rule scores and ranks all eligible activity--mode pairs. Only the highest-ranked mode for each activity is retained, and knee-point selection identifies a promising subset for further consideration.
    \item \textbf{Activity-group selection}: Combinations of the promising pairs are enumerated and resource-infeasible groups are removed. A group selection rule scores the remaining groups and selects one group for concurrent execution.
\end{enumerate}

The selected activities are started, their realized durations are sampled, and their resource demands are reserved. When no additional activity can start at the current time, the simulation clock advances to the next activity-completion event. The state and eligible set are then updated and the procedure repeats until every activity is completed. Consequently, ordering-rule evaluation, feasible-group enumeration, group-rule evaluation, and resource-feasibility checks are repeatedly executed within a single simulation, making this loop a major contributor to total runtime.

\subsection{Evolving Heuristic Rules via Genetic Programming}

\begin{figure}[h]
\centering
\includegraphics[width=\columnwidth]{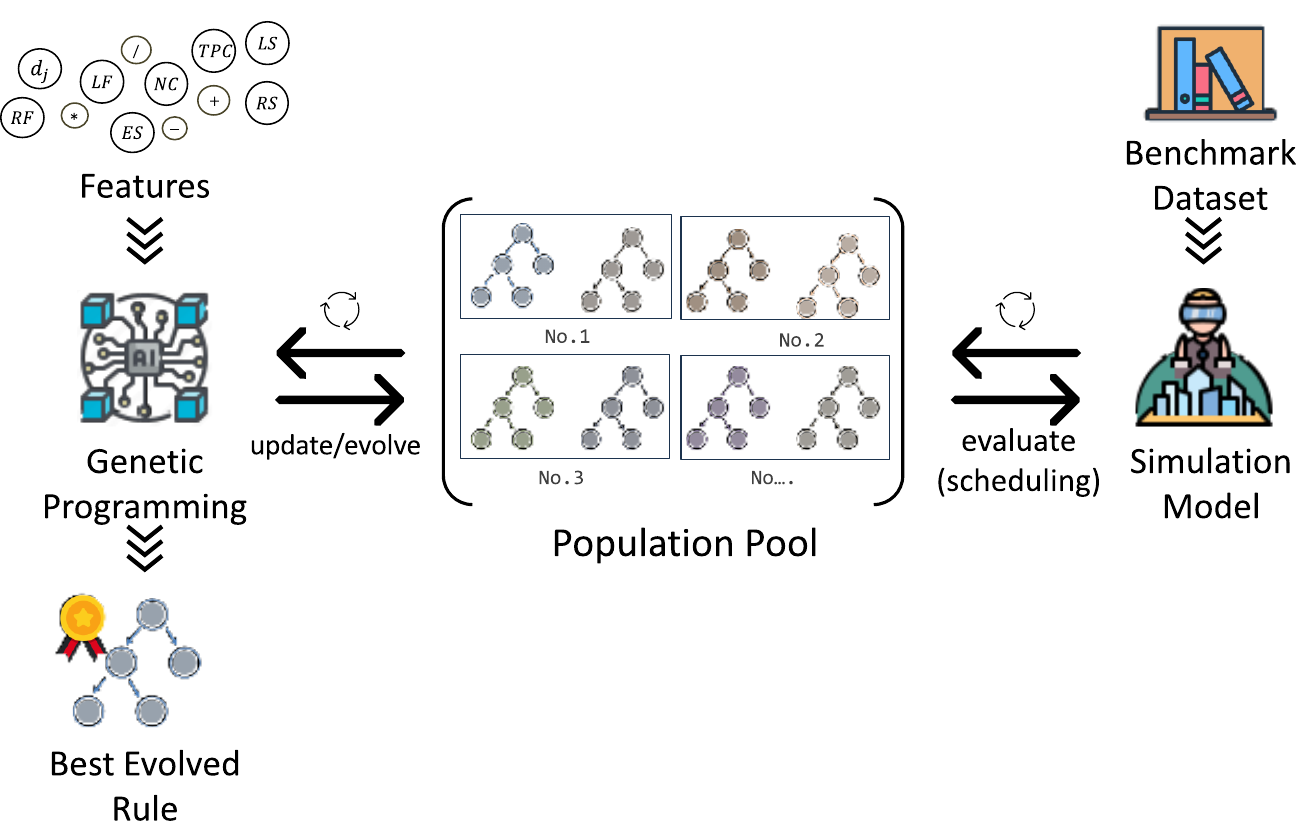}
\caption{Genetic Programming Hyper-Heuristic (GPHH) framework: candidate priority rules are represented as expression trees, evolved against a benchmark dataset, and evaluated via the simulation model.}
\label{fig:gphh}
\end{figure}

To automate rule design, this project implements a Genetic Programming Hyper-Heuristic (GPHH) framework \cite{nguyenGeneticProgrammingProduction2017,zhangSurveyGeneticProgramming2023} that evolves the two priority rules simultaneously. Candidate rule pairs are modeled as mathematical expression trees using simulation state variables (terminal sets) and operators (function sets), as shown in Fig.~\ref{fig:gphh}. During training, each pair directs all scheduling and mode choices within the Python simulator across a suite of DMRCPSP instances. The resulting makespan determines evolutionary fitness. Because the outer GPHH loop evaluates thousands of trees over millions of nested simulator iterations, internal Python interpretive overhead compounds exponentially, causing severe cluster resource exhaustion.

\begin{figure*}[t]
\centering
\includegraphics[width=\textwidth]{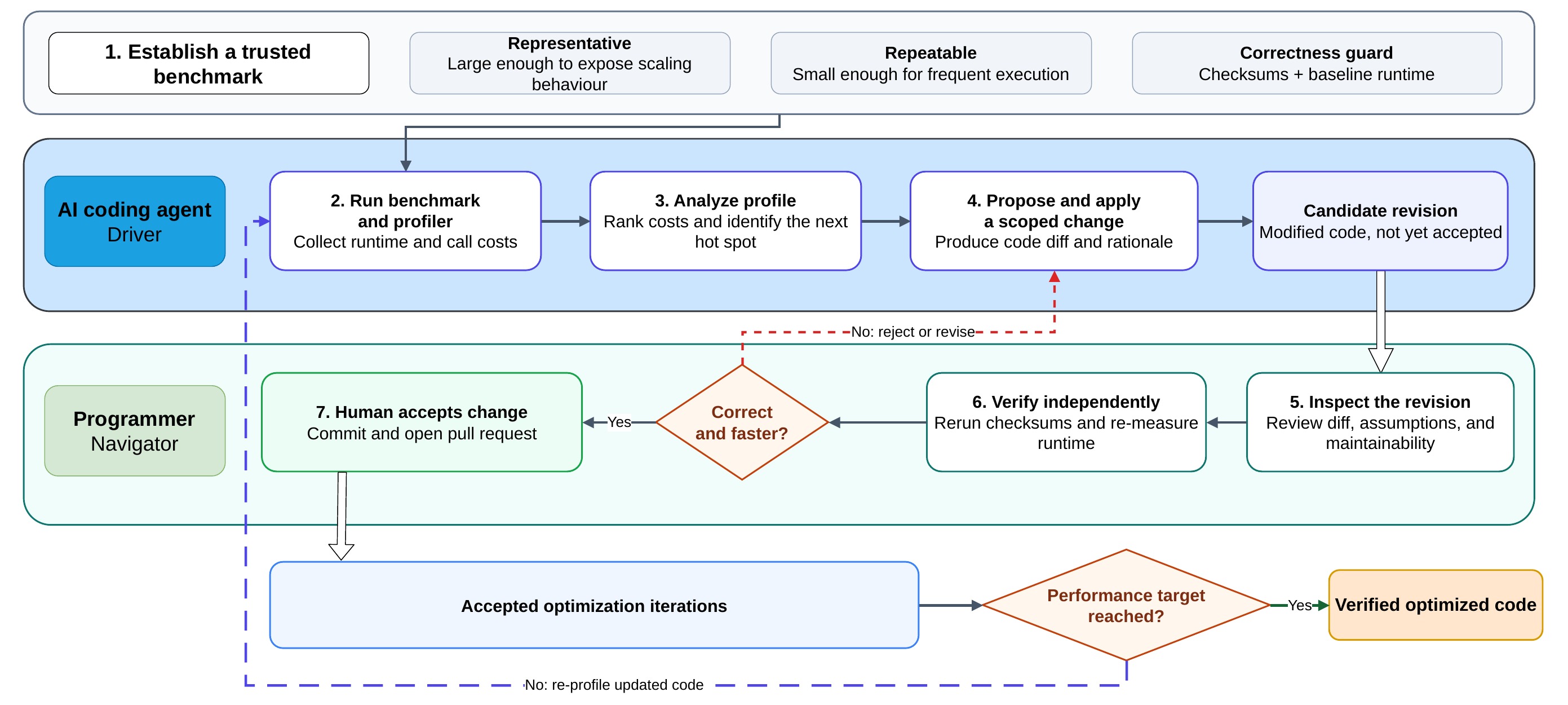}
\caption{Semi-automated, human-in-the-loop code-optimization workflow. The AI coding agent acts as the driver by executing the benchmark and profiler, identifying the next hot spot, and applying a scoped change. The programmer acts as the navigator by inspecting the revision, independently validating checksums and runtime, and retaining final authority over commits and pull requests. Accepted changes are re-profiled iteratively until the performance target is reached.}
\label{fig:agentic-optimization-workflow}
\end{figure*}

\section{Implementation \& Optimization Strategies} \label{sec:optimization}
Figure~\ref{fig:agentic-optimization-workflow} summarizes the implementation and optimization workflow used in this case study. The first step to improving code performance involves generating a test case, which should be small enough to allow the code to be run many times, yet large enough to capture the large execution time asymptotic behavior of the code. The test should also contain check sums (numbers that are representative of the solution) to ensure that changes do not break the code.

Next, performance improvement efforts should involve profiling, i.e., identifying parts of a program where most of the execution time is spent. Finding execution bottlenecks allows one to focus attention on parts that are known to contribute most to the runtime. The process is iterative: as each hot spot is addressed, other bottlenecks emerge and need to be suppressed until satisfactory performance is achieved.

These two steps lend themselves naturally to automation by an AI coding agent -- Claude \cite{anthropic_claude_2026}. The agent can run the test, analyze the profiling output, identify the next hot spot to address, and implement the corresponding code change. In this work we instead adopted a semi-automated workflow, in which the agent proposes and applies code modifications while the programmer inspects the output and independently re-measures execution time to confirm that each change is both correct and beneficial. This division of labor mirrors the pair-programming practice advocated by Extreme Programming \cite{beck2000extreme,williams2000strengthening}, with the agent taking the role of the ``driver'' and the programmer that of the ``navigator'' -- except that the driver is a machine. Responsibility for committing changes and opening pull requests remains with the human programmer, who retains final authority over what enters the code base.

We categorized the performance engineering efforts into four core strategies, ranking them by algorithmic complexity and resource savings.

\subsection{Environmental Runtime Cleansing (Baseline Cleanup)}

The first optimization phase focused on eliminating avoidable execution overhead introduced by development-oriented instrumentation embedded within the production workflow. The original implementation had evolved through multiple development cycles and retained several diagnostic mechanisms—such as debug-level string evaluations, runtime validation checks, and excessive logging operations within the simulation loop—that were valuable during algorithm development but highly inefficient during large-scale HPC execution. Because the genetic programming hyper-heuristic framework evaluates large populations of candidate solutions through iterative simulation, these individually inexpensive operations were repeatedly triggered across millions of scheduling state evaluations. When aggregated across thousands of evolutionary generations and repeated scheduling simulations, these small, per-operation overheads within the inner execution loops created massive cumulative computational overhead, directly driving up total HPC resource consumption.

\paragraph{Action}
The first optimization step applied a non-invasive runtime cleansing procedure designed to remove unnecessary production overhead without altering algorithmic behavior or modifying the underlying scheduling logic. The following changes were applied:

\begin{itemize}
    \item \textbf{Disabled debug-level execution paths.} Development-only diagnostic blocks, including verbose logging statements, intermediate state reporting, and runtime inspection routines, were deactivated during production execution. These operations were previously executed within high-frequency simulation pathways and contributed unnecessary CPU cycles and memory operations.
    \item \textbf{Enabled Python optimized execution mode.} The application was executed using Python's optimized runtime flag:
\begin{lstlisting}
python -O script.py
\end{lstlisting}
    This instructs the Python interpreter to enable optimization mode by:
    \begin{itemize}
        \item removing execution of \texttt{assert} statements,
        \item suppressing generation of debugging metadata,
        \item reducing interpreter-level validation overhead,
        \item generating optimized bytecode (\texttt{.opt-1.pyc}).
    \end{itemize}
    Importantly, this modification required no changes to the algorithm implementation, data structures, or scheduling model, allowing the impact of environmental optimization to be isolated independently from later code-level improvements.
    \item \textbf{Preserved computational equivalence.} Since the optimization only removed non-functional instrumentation, the generated scheduling solutions and evolutionary search behavior remained unchanged. This ensured that runtime improvements represented genuine computational efficiency gains rather than algorithmic approximation or reduced solution quality.
\end{itemize}

\paragraph{Impact}
Environmental runtime cleansing provided an immediate improvement in execution efficiency, reducing total simulation runtime from 1,298 seconds to 787 seconds, representing a reduction of 511 seconds (8.5 minutes) per execution. This corresponds to an approximately 40\% reduction in runtime without any structural modification to the codebase.

\subsection{Low-Complexity Restructuring: Static Dictionary Caching}

After removing runtime instrumentation overhead, the next optimization targeted the dominant computational bottleneck identified through profiling: repeated graph queries inside the dynamic Critical Path Method (CPM) calculation. The genetic programming hyper-heuristic simulator repeatedly evaluates candidate scheduling decisions by recomputing precedence relationships among unfinished activities. This dynamic CPM calculation is executed thousands of times during evolutionary search and therefore represents a performance-critical execution pathway.

The original implementation relied on the NetworkX \cite{hagberg2008} graph abstraction to retrieve predecessor and successor relationships during each simulation step. Although NetworkX provides a convenient and flexible interface for graph manipulation, its dynamic graph views introduce substantial overhead when repeatedly accessed inside deeply nested simulation loops. Each query requires traversal through Python-level graph objects, including adjacency dictionaries wrapped by NetworkX view objects, dynamic attribute resolution, and iterator creation.

Profiling identified these NetworkX operations as a significant contributor to runtime. In particular, operations such as:

\begin{lstlisting}
_subgraph.predecessors(cur_act)
_subgraph.successors(cur_act)
graph_nx.successors(task)
\end{lstlisting}

were repeatedly executed during the forward and backward passes of the CPM calculation.

Because the project dependency graph remains static throughout the simulation process, repeatedly querying the NetworkX representation provided no computational benefit. The graph topology does not change during genetic programming evaluation; only the scheduling state changes.

\paragraph{Action}
The optimization replaced dynamic NetworkX graph access with pre-computed native Python dictionary structures. During problem initialization, the graph topology was extracted into direct adjacency mappings:

\begin{lstlisting}
predecessors_dict[node] = {predecessor_nodes}
neighbors_dict[node] = {successor_nodes}
\end{lstlisting}

These structures were then accessed directly during simulation:

\begin{lstlisting}
pred_dict = self.rcpsp_problem.graph.predecessors_dict
succ_dict = self.rcpsp_problem.graph.neighbors_dict
\end{lstlisting}

The CPM forward pass was modified from dynamic NetworkX traversal:

\begin{lstlisting}
pred_acts = list(_subgraph.predecessors(cur_act))
\end{lstlisting}

to direct dictionary lookup:

\begin{lstlisting}
pred_acts = [
    p for p in pred_dict.get(cur_act, set())
    if p in subproblem_set
]
\end{lstlisting}

Similarly, backward CPM propagation was changed from:

\begin{lstlisting}
succ_acts = list(_subgraph.successors(cur_act))
\end{lstlisting}

to:

\begin{lstlisting}
succ_acts = [
    s for s in succ_dict.get(cur_act, set())
    if s in subproblem_set
]
\end{lstlisting}

Additional restructuring included: replacing NetworkX subgraph construction with direct set-based node filtering; avoiding repeated graph object creation; and replacing graph-size queries such as

\begin{lstlisting}
_subgraph.number_of_nodes()
\end{lstlisting}

with direct collection size evaluation:

\begin{lstlisting}
len(_subproblem)
\end{lstlisting}

preserving NetworkX only for offline graph construction and analysis rather than runtime simulation.

The modification required limited code changes because the optimization preserved the original algorithmic logic. The CPM calculation, precedence constraints, and scheduling decisions remained unchanged; only the underlying graph access mechanism was replaced.

\paragraph{Impact}
This low-complexity restructuring reduced runtime by 273 seconds (4.5 minutes), representing approximately 30\% performance improvement for the simulator execution workload. The improvement was achieved without changing the genetic programming representation, modifying heuristic operators, altering scheduling constraints, reducing simulation accuracy, or changing solution quality. Instead, the optimization removed unnecessary abstraction overhead by moving invariant graph processing from the simulation phase into the initialization phase.

This result demonstrates an important performance engineering principle for scientific Python workloads: general-purpose data structures should not remain inside high-frequency computational kernels when their underlying information is static. By replacing dynamic graph views with lightweight adjacency tables, the framework retained NetworkX's usability during model construction while achieving significantly improved execution efficiency during large-scale evolutionary optimization.

\subsection{Algorithmic Restructuring: Incremental Resource Feasibility Tracking}

After optimizing graph access and simulation-level overhead, profiling identified the feasibility checking routine (\texttt{check\_feasible}) as another major runtime bottleneck. This routine is responsible for evaluating candidate activity combinations during the activity group selection process of the scheduling simulator.

The original implementation used a branch-and-bound enumeration strategy to identify feasible activity groups under resource constraints. At each node of the search tree, the algorithm considered extending the current partial solution by adding a new activity. Before accepting the extension, the feasibility checker recalculated the total resource consumption of the complete partial combination. For every candidate expansion, the original implementation performed a full aggregation operation:

\begin{lstlisting}
resource_consumed = np.sum(
    [
        self.rcpsp_problem.mode_resources_matrix[comb[0]][comb[1]]
        for comb in new_combination
    ],
    axis=0,
)
\end{lstlisting}

This required iterating through all previously selected activities in the current combination and summing their resource requirements again, even though most of the resource consumption state remained unchanged between neighboring search-tree nodes.

Because the branch-and-bound procedure explores a large number of candidate combinations, the same resource contributions were repeatedly recomputed across many related search states. The computational cost therefore grew unnecessarily with the depth and breadth of the enumeration tree.

\paragraph{Action}
A structural algorithmic redesign was introduced by replacing full recomputation with incremental resource state propagation. Instead of deriving the resource usage from the complete activity combination at every search step, the current resource consumption vector was treated as part of the search state and propagated through the branch-and-bound tree. The revised implementation introduced a persistent resource tracking variable:

\begin{lstlisting}
resource_consumed
\end{lstlisting}

which represents the accumulated resource demand of the current partial combination. When a new activity is considered, only the incremental resource contribution of that activity is calculated:

\begin{lstlisting}
new_resource_consumed = (
    resource_consumed
    + self.rcpsp_problem.mode_resources_matrix[new_item[0]][new_item[1]]
)
\end{lstlisting}

Feasibility is then evaluated directly against the available resources:

\begin{lstlisting}
bool(np.all(current_resource_avail >= new_resource_consumed))
\end{lstlisting}

The search state stored in the priority queue was extended from:

\begin{lstlisting}
(priority, id, combination, remaining)
\end{lstlisting}

to:

\begin{lstlisting}
(priority, id, combination, remaining, resource_consumed)
\end{lstlisting}

This allowed each child node in the search tree to inherit the resource state of its parent and update only the newly introduced activity demand.

\paragraph{Additional Structural Improvements}
The optimization also reduced repeated computation by moving invariant operations outside the search loop.

\emph{1) Resource availability calculation moved outside feasibility checking:} Previously, each feasibility evaluation accessed

\begin{lstlisting}
self.resource_avail_in_time[:, self.current_time]
\end{lstlisting}

inside the checking function. This was replaced with a single calculation before branch-and-bound expansion:

\begin{lstlisting}
current_resource_avail = (
    self.resource_avail_in_time[:, self.current_time]
)
\end{lstlisting}

Since resource availability does not change during a single activity-group selection procedure, repeated retrieval was unnecessary.

\emph{2) Initial resource state construction:} The search tree was initialized with an empty resource vector:

\begin{lstlisting}
_zero_resources = np.zeros(
    self.rcpsp_problem.mode_resources_matrix[
        eligibles[0][0]
    ][eligibles[0][1]].shape,
    dtype=float,
)
\end{lstlisting}

The initial heap state was then extended:

\begin{lstlisting}
(0, max_id, [], list(eligibles), _zero_resources)
\end{lstlisting}

allowing resource information to flow naturally through the search process.

\paragraph{Impact}
The incremental resource tracking architecture reduced execution time by 330 seconds (5.5 minutes) compared with the previous implementation. Unlike previous optimizations that primarily removed software overhead, this improvement required an algorithmic restructuring of the search procedure. The optimization achieved performance gains by reducing unnecessary repeated computation while preserving the exact feasibility evaluation logic. The improvement provides several advantages:

\begin{itemize}
    \item \textbf{Reduced computational complexity:} resource feasibility checks changed from repeated summation over partial solutions to constant-time incremental updates.
    \item \textbf{Improved scalability:} deeper branch-and-bound exploration benefits more significantly because each additional search level avoids recomputing accumulated resource usage.
    \item \textbf{Preserved solution quality:} the branch-and-bound search space and feasibility constraints remain unchanged.
    \item \textbf{Maintained deterministic behavior:} candidate selection and scheduling decisions are identical before and after optimization.
\end{itemize}

This optimization demonstrates an important principle in evolutionary scheduling systems: state information that evolves predictably during search should be maintained incrementally rather than reconstructed repeatedly. By treating resource consumption as part of the search state, the simulator substantially reduced computational overhead within one of its most frequently executed decision-making components.

\subsection{Medium-Complexity Compilation: Accelerating Critical Scheduling Kernels with Numba JIT}

After addressing high-level architectural inefficiencies, the remaining performance bottlenecks were concentrated in computational kernels that performed repeated feasibility searches during schedule generation. One such hot spot was the \texttt{earliestStart} function, which determines the earliest feasible start time for an activity by scanning the available resource timeline.

The \texttt{earliestStart} operation is executed extensively during simulation because every candidate scheduling decision requires evaluating whether an activity can be placed within the current resource constraints. For large resource-constrained project scheduling problems (RCPSP), this function may be invoked millions of times during genetic programming evaluation.

The original implementation used a pure Python and NumPy-based approach:

\begin{lstlisting}
result = self.resource_avail_in_time - resource_req

for t in range(from_time, result.shape[1] - duration + 1):
    if np.all(result[:, t:t+duration] >= 0):
        return t
\end{lstlisting}

Although NumPy provides efficient vectorized operations, this implementation introduced several hidden costs when executed at high frequency:

\begin{itemize}
    \item \textbf{Repeated temporary array allocation.} The subtraction operation \texttt{resource\_avail\_in\_time - resource\_req} generated a new intermediate array for every invocation. For millions of calls, these temporary allocations created significant memory traffic and garbage collection overhead.
    \item \textbf{Repeated slicing operations.} Each candidate time window required creation of a NumPy slice \texttt{result[:, t:t+duration]}, which introduced additional indexing overhead.
    \item \textbf{Python-level iteration around vector operations.} Although the inner comparison was performed by NumPy, the outer search loop remained controlled by Python. The repeated transition between Python execution and optimized NumPy routines limited performance gains.
\end{itemize}

Profiling showed that this function represented a significant fraction of simulator execution time, making it an ideal candidate for low-level compilation.

\paragraph{Action}
The \texttt{earliestStart} computation was redesigned as a compiled numerical kernel using Numba Just-In-Time (JIT) \cite{lamNumbaLLVMbasedPython2015} compilation. A dedicated Numba-compatible implementation was introduced:

\begin{lstlisting}
@numba.njit(cache=True)
def _earliest_start_nb(
    resource_avail: np.ndarray,
    resource_req: np.ndarray,
    from_time: int,
    duration: int,
) -> int:
\end{lstlisting}

The function was rewritten using explicit loops over candidate start times, activity duration windows, and resource dimensions. The search algorithm directly examines resource feasibility:

\begin{lstlisting}
for t in range(from_time, end):
    ok = True
    for s in range(t, t + duration):
        for r in range(n_resources):
            if resource_avail[r, s] < resource_req[r]:
                ok = False
                break
\end{lstlisting}

This design enables LLVM-based compilation by Numba because all operations operate on primitive integer values, contiguous NumPy arrays, and statically typed numerical operations. The original object-heavy implementation was replaced with a lightweight numerical kernel.

\paragraph{Runtime Optimization Strategies}
The compiled implementation introduced several additional improvements.

\emph{1) Elimination of temporary array creation:} Instead of computing \texttt{resource\_avail - resource\_req} for every invocation, the kernel directly compares resource availability:

\begin{lstlisting}
resource_avail[r, s] < resource_req[r]
\end{lstlisting}

This avoids allocating intermediate arrays. For workloads containing millions of scheduling evaluations, avoiding these allocations significantly reduces memory movement and interpreter overhead.

\emph{2) Early termination during infeasible checks:} The original implementation evaluated complete NumPy expressions before determining feasibility. The Numba kernel introduces early exit behaviour:

\begin{lstlisting}
if resource_avail[r, s] < resource_req[r]:
    ok = False
    break
\end{lstlisting}

As soon as a single resource constraint violation is identified, the remaining checks for that candidate window are skipped. This is particularly effective because many candidate scheduling positions are infeasible and can be rejected early.

\emph{3) Removal of Python interpreter overhead:} The compiled function executes the nested search loops as optimized machine code through LLVM. This eliminates Python loop interpretation, repeated function dispatch, NumPy temporary object management, and dynamic type checking. The Python-facing simulator now only performs the initial data preparation and receives the computed start time:

\begin{lstlisting}
t = _earliest_start_nb(
    self.resource_avail_in_time,
    resource_req,
    from_time,
    duration,
)
\end{lstlisting}

\paragraph{Impact}
The Numba-based compilation reduced total runtime by 184 seconds (approximately 3 minutes) compared with the previous optimized implementation. The improvement was achieved with medium implementation complexity because the optimization required identifying a suitable numerical hot spot, separating the computational kernel from object-oriented simulator code, converting data structures into Numba-compatible NumPy arrays, and replacing high-level vector operations with explicit compiled loops. However, the scheduling algorithm itself remained unchanged. The optimization preserved identical resource feasibility constraints, identical earliest start decisions, and identical generated schedules.

The improvement demonstrates that carefully selected JIT compilation can provide substantial performance benefits for scientific Python applications without requiring migration to a lower-level programming language.

\section{Evaluation \& Financial Impact}

The combined application of these architectural refactoring tiers was tested across production datasets on the REANNZ HPC network infrastructure. Fig. \ref{tab_summary} summarizes the runtime savings, implementation effort, and primary system benefit of each optimization layer.

\begin{figure}[htbp]
\centering
\begin{tabular}{p{2.0cm}p{1.1cm}p{1.1cm}p{3.2cm}}
\toprule
\textbf{Optimization Layer} & \textbf{Time Saved} & \textbf{Effort} & \textbf{Primary Benefit} \\
\midrule
Debug flag overhaul (\texttt{-O}) & 511 s & Minimal & Eradication of logging \& assertion overhead \\
\midrule
Pre-built dict over NetworkX views & 273 s & Low & Elimination of dynamic library lookups \\
\midrule
Incremental resource tracking & 330 s & High & Asymptotic reduction in calculation loops \\
\midrule
Numba-JIT \texttt{earliestStart} & 184 s & Medium & Native loop compilation via LLVM primitives \\
\bottomrule
\end{tabular}
\caption{Summary of Optimization Layers and Their Impact: Runtime reduction achieved by the four optimization layers, from the legacy baseline (1298 seconds, 5.17M compute units/yr) to the optimized implementation (195 seconds, 1.05M compute units/yr).}
\label{tab_summary}
\end{figure}

The cumulative impact of the refactoring pipeline dropped total simulation time from a non-optimized benchmark of 1,298 seconds to 195 seconds, an 85\% execution compute time saved on the HPC. Fig. \ref{tab_summary} illustrates the cumulative effect of each optimization layer on total runtime and annual compute unit consumption.

Prior to this consultancy, the project's historical footprint of \textasciitilde5,173,000 compute units violated the allocated ceiling. By compressing execution cycles, the optimized code now fits comfortably within the newly approved 1,054,000 compute unit quota. The refactored framework saves over 4 million compute units (equivalent to CPU core-hours) per year, which translates into an annual reduction of NZ~\$320,000 (approximately US \$188,000) in compute costs.

\section{Conclusions}

This work demonstrates a practical AI-assisted workflow that treats scientific code optimization as a continuous and evidence-driven process. Agentic AI can help researchers identify performance bottlenecks, propose and evaluate improvements, and provide explanations for code changes. This makes performance engineering more accessible to domain researchers with limited specialist software-engineering expertise, while human review preserves scientific correctness and control over the codebase.

The workflow has potential applications beyond the project-scheduling problem studied here, particularly for simulation-intensive scheduling and other experiment-intensive research domains. Applying this approach to performance-limited research software could improve HPC resource efficiency, reduce computational costs, and accelerate scientific discovery. Future work will investigate its application across a broader range of research projects where software performance limits experimental scale and productivity.

\bibliographystyle{IEEEtran}
\bibliography{biblio}

@article{pellAmbulanceResponse2001,
  author  = {Pell, Jill P. and Sirel, Jane M. and Marsden, Andrew K. and Ford, Ian and Cobbe, Stuart M.},
  title   = {Effect of Reducing Ambulance Response Times on Deaths from Out of Hospital Cardiac Arrest: Cohort Study},
  journal = {BMJ},
  year    = {2001},
  volume  = {322},
  number  = {7299},
  pages   = {1385--1388},
  doi     = {10.1136/bmj.322.7299.1385},
  url     = {https://doi.org/10.1136/bmj.322.7299.1385}
}

@article{schmidDynamicAmbulance2012,
  author  = {Schmid, Verena},
  title   = {Solving the Dynamic Ambulance Relocation and Dispatching Problem Using Approximate Dynamic Programming},
  journal = {European Journal of Operational Research},
  year    = {2012},
  volume  = {219},
  number  = {3},
  pages   = {611--621},
  doi     = {10.1016/j.ejor.2011.10.043},
  url     = {https://doi.org/10.1016/j.ejor.2011.10.043}
}

@inproceedings{maclachlanAmbulanceSubset2022,
  author    = {MacLachlan, Jordan and Mei, Yi and Zhang, Fangfang and Zhang, Mengjie},
  title     = {Genetic Programming for Vehicle Subset Selection in Ambulance Dispatching},
  booktitle = {2022 IEEE Congress on Evolutionary Computation (CEC)},
  year      = {2022},
  pages     = {1--8},
  publisher = {IEEE},
  doi       = {10.1109/CEC55065.2022.9870323},
  url       = {https://doi.org/10.1109/CEC55065.2022.9870323}
}

@inproceedings{maclachlanEmergencyDispatch2023,
  author    = {MacLachlan, Jordan and Mei, Yi and Zhang, Fangfang and Zhang, Mengjie and Signal, Jessica},
  title     = {Learning Emergency Medical Dispatch Policies Via Genetic Programming},
  booktitle = {Proceedings of the Genetic and Evolutionary Computation Conference},
  year      = {2023},
  pages     = {1409--1417},
  publisher = {Association for Computing Machinery},
  address   = {New York, NY, USA},
  doi       = {10.1145/3583131.3590434},
  url       = {https://doi.org/10.1145/3583131.3590434}
}

@article{hulshofHealthcarePlanning2012,
  author  = {Hulshof, Peter J. H. and Kortbeek, Nikky and Boucherie, Richard J. and Hans, Erwin W. and Bakker, Piet J. M.},
  title   = {Taxonomic Classification of Planning Decisions in Health Care: A Structured Review of the State of the Art in {OR/MS}},
  journal = {Health Systems},
  year    = {2012},
  volume  = {1},
  number  = {2},
  pages   = {129--175},
  doi     = {10.1057/hs.2012.18},
  url     = {https://doi.org/10.1057/hs.2012.18}
}

@article{zhaoRealtimeJobShop2022,
  author  = {Zhao, Anran and Liu, Peng and Gao, Xiyu and Huang, Guotai and Yang, Xiuguang and Ma, Yuan and Xie, Zheyu and Li, Yunfeng},
  title   = {Data-Mining-Based Real-Time Optimization of the Job Shop Scheduling Problem},
  journal = {Mathematics},
  year    = {2022},
  volume  = {10},
  number  = {23},
  pages   = {4608},
  doi     = {10.3390/math10234608},
  url     = {https://doi.org/10.3390/math10234608}
}

@article{luSmartGridScheduling2018,
  author  = {Lu, Xinhui and Zhou, Kaile and Zhang, Xiaoling and Yang, Shanlin},
  title   = {A Systematic Review of Supply and Demand Side Optimal Load Scheduling in a Smart Grid Environment},
  journal = {Journal of Cleaner Production},
  year    = {2018},
  volume  = {203},
  pages   = {757--768},
  doi     = {10.1016/j.jclepro.2018.08.301},
  url     = {https://doi.org/10.1016/j.jclepro.2018.08.301}
}

@article{blazewiczSchedulingComplexity1983,
  author  = {B{\l}a{\.z}ewicz, Jacek and Lenstra, Jan Karel and Rinnooy Kan, Alexander H. G.},
  title   = {Scheduling Subject to Resource Constraints: Classification and Complexity},
  journal = {Discrete Applied Mathematics},
  year    = {1983},
  volume  = {5},
  number  = {1},
  pages   = {11--24},
  doi     = {10.1016/0166-218X(83)90012-4},
  url     = {https://doi.org/10.1016/0166-218X(83)90012-4}
}

@article{hartmannCompetitiveGA1998,
  author  = {Hartmann, S{\"o}nke},
  title   = {A Competitive Genetic Algorithm for Resource-Constrained Project Scheduling},
  journal = {Naval Research Logistics},
  year    = {1998},
  volume  = {45},
  number  = {7},
  pages   = {733--750},
  doi     = {10.1002/(SICI)1520-6750(199810)45:7<733::AID-NAV5>3.0.CO;2-C},
  url     = {https://doi.org/10.1002/(SICI)1520-6750(199810)45:7<733::AID-NAV5>3.0.CO;2-C}
}

@article{vallsHybridGA2008,
  author  = {Valls, Vicente and Ballest{\'i}n, Francisco and Quintanilla, Sacramento},
  title   = {A Hybrid Genetic Algorithm for the Resource-Constrained Project Scheduling Problem},
  journal = {European Journal of Operational Research},
  year    = {2008},
  volume  = {185},
  number  = {2},
  pages   = {495--508},
  doi     = {10.1016/j.ejor.2006.12.033},
  url     = {https://doi.org/10.1016/j.ejor.2006.12.033}
}

@article{amaranSimulationOptimization2016,
  author  = {Amaran, Satyajith and Sahinidis, Nikolaos V. and Sharda, Bikram and Bury, Scott J.},
  title   = {Simulation Optimization: A Review of Algorithms and Applications},
  journal = {Annals of Operations Research},
  year    = {2016},
  volume  = {240},
  number  = {1},
  pages   = {351--380},
  doi     = {10.1007/s10479-015-2019-x},
  url     = {https://doi.org/10.1007/s10479-015-2019-x}
}

@article{fujimotoParallelSimulation2016,
  author  = {Fujimoto, Richard M.},
  title   = {Research Challenges in Parallel and Distributed Simulation},
  journal = {ACM Transactions on Modeling and Computer Simulation},
  year    = {2016},
  volume  = {26},
  number  = {4},
  pages   = {1--29},
  doi     = {10.1145/2866577},
  url     = {https://doi.org/10.1145/2866577}
}

@article{tangGPUDiscreteEvent2013,
  author  = {Tang, Wenjie and Yao, Yiping},
  title   = {A {GPU}-Based Discrete Event Simulation Kernel},
  journal = {SIMULATION},
  year    = {2013},
  volume  = {89},
  number  = {11},
  pages   = {1335--1354},
  doi     = {10.1177/0037549713508839},
  url     = {https://doi.org/10.1177/0037549713508839}
}

@inproceedings{paszke2017automatic,
  title={Automatic differentiation in PyTorch},
  author={Paszke, Adam and Gross, Sam and Chintala, Soumith and Chanan, Gregory and Yang, Edward and DeVito, Zachary and Lin, Zeming and Desmaison, Alban and Antiga, Luca and Lerer, Adam},
  booktitle={NIPS-W},
  year={2017},
  url={https://openreview.net/pdf/25b8eee6c373d48b84e5e9c6e10e7cbbbce4ac73.pdf}
}

@Article{harris2020array,
title = {Array programming with {NumPy}},
author = {Charles R. Harris and K. Jarrod Millman and St{\'e}fan J. van der Walt and Ralf Gommers and Pauli Virtanen and others},
year = {2020},
journal = {Nature},
volume = {585},
pages = {357--362},
doi = {10.1038/s41586-020-2649-2},
url = {https://doi.org/10.1038/s41586-020-2649-2}
}

@misc{reback2020pandas,
    author       = {{The pandas development team}},
    title        = {pandas-dev/pandas: Pandas},
    month        = feb,
    year         = 2020,
    publisher    = {Zenodo},
    version      = {latest},
    doi          = {10.5281/zenodo.3509134},
    url          = {https://doi.org/10.5281/zenodo.3509134}
}

@article{williams2000strengthening,
  author  = {Laurie Williams and Robert R. Kessler and Ward Cunningham and Ron Jeffries},
  title   = {Strengthening the Case for Pair Programming},
  journal = {IEEE Software},
  volume  = {17},
  number  = {4},
  pages   = {19--25},
  year    = {2000},
  doi     = {10.1109/52.854064},
  url     = {https://doi.org/10.1109/52.854064}
}

@book{beck2000extreme,
  author    = {Kent Beck},
  title     = {Extreme Programming Explained: Embrace Change},
  publisher = {Addison-Wesley},
  address   = {Reading, MA},
  year      = {2000},
  isbn      = {9780201616415},
  url       = {https://books.google.com/books?id=G8EL4H4vf7UC}
}

@article{christodoulouResourceConstrained2012,
  author  = {Christodoulou, Symeon E. and Tezias, Elias S. and Galaras, Kyriakos A.},
  title   = {Resource-Constrained Scheduling of Construction Projects and Simulation of the Entropy Impact on a Project's Duration and Cost},
  journal = {International Journal of Project Organisation and Management},
  year    = {2012},
  volume  = {4},
  number  = {4},
  pages   = {322--338},
  doi     = {10.1504/IJPOM.2012.050328},
  url     = {https://doi.org/10.1504/IJPOM.2012.050328}
}

@article{zhangPythonInterpretation2022,
  author  = {Zhang, Qiang and Xu, Lei and Zhang, Xiangyu and Xu, Baowen},
  title   = {Quantifying the Interpretation Overhead of Python},
  journal = {Science of Computer Programming},
  year    = {2022},
  volume  = {215},
  pages   = {102759},
  doi     = {10.1016/j.scico.2021.102759},
  url     = {https://doi.org/10.1016/j.scico.2021.102759}
}

@incollection{tianScalableKneePoint2026,
  author    = {Tian, Yuan and Mei, Yi and Zhang, Mengjie},
  title     = {Scalable Knee-Point Guided Activity Group Selection in Multi-Tree Genetic Programming for Dynamic Multi-Mode Project Scheduling},
  booktitle = {PRICAI 2025: Trends in Artificial Intelligence},
  editor    = {Mei, Yi and Qian, Chao and Bai, Quan and Xue, Bing and Khanna, Sankalp},
  year      = {2026},
  volume    = {16454},
  pages     = {576--592},
  publisher = {Springer Nature Singapore},
  address   = {Singapore},
  doi       = {10.1007/978-981-95-7081-2_40},
  url       = {https://doi.org/10.1007/978-981-95-7081-2_40}
}

@article{bold2022faster,
  author  = {Bold, Matthew and Goerigk, Marc},
  title   = {A faster exact method for solving the robust multi-mode resource-constrained project scheduling problem},
  journal = {Operations Research Letters},
  volume  = {50},
  number  = {5},
  pages   = {581--587},
  year    = {2022},
  doi     = {10.1016/j.orl.2022.08.003},
  note    = {Preprint: arXiv:2203.06983},
  url     = {https://doi.org/10.1016/j.orl.2022.08.003}
}

@article{peteghemExperimentalInvestigationMetaheuristics2014,
  author  = {Van Peteghem, Vincent and Vanhoucke, Mario},
  title   = {An Experimental Investigation of Metaheuristics for the Multi-Mode Resource-Constrained Project Scheduling Problem on New Dataset Instances},
  journal = {European Journal of Operational Research},
  year    = {2014},
  volume  = {235},
  number  = {1},
  pages   = {62--72},
  doi     = {10.1016/j.ejor.2013.10.012},
  url     = {https://doi.org/10.1016/j.ejor.2013.10.012}
}

@article{hartmannUpdatedSurveyVariants2022,
  author  = {Hartmann, S{\"o}nke and Briskorn, Dirk},
  title   = {An Updated Survey of Variants and Extensions of the Resource-Constrained Project Scheduling Problem},
  journal = {European Journal of Operational Research},
  year    = {2022},
  volume  = {297},
  number  = {1},
  pages   = {1--14},
  doi     = {10.1016/j.ejor.2021.05.004},
  url     = {https://doi.org/10.1016/j.ejor.2021.05.004}
}

@article{asadujjamanSupplyChainIntegrated2024,
  author  = {Asadujjaman, Md. and Rahman, Humyun Fuad and Chakrabortty, Ripon K. and Ryan, Michael J.},
  title   = {Supply Chain Integrated Resource-Constrained Multi-Project Scheduling Problem},
  journal = {Computers \& Industrial Engineering},
  year    = {2024},
  volume  = {194},
  pages   = {110380},
  doi     = {10.1016/j.cie.2024.110380},
  url     = {https://doi.org/10.1016/j.cie.2024.110380}
}

@article{chenResourceconstrainedProjectScheduling2024,
  author  = {Chen, Gang and He, Wen and Tian, Yu and Ma, Ke},
  title   = {Resource-Constrained Project Scheduling with Multiple States: Bi-Objective Optimization Model and Case Study of Aircraft Maintenance},
  journal = {Computers \& Industrial Engineering},
  year    = {2024},
  volume  = {191},
  pages   = {110169},
  doi     = {10.1016/j.cie.2024.110169},
  url     = {https://doi.org/10.1016/j.cie.2024.110169}
}

@inproceedings{kurtResourceConstrainedMultiproject2018,
  author    = {Kurt, Pelin Akyil and Kececi, Baris},
  title     = {Resource Constrained Multi-Project Scheduling: Application in Software Company},
  booktitle = {Advances in Manufacturing, Production Management and Process Control},
  year      = {2018},
  pages     = {549--557},
  publisher = {Springer International Publishing},
  address   = {Cham},
  doi       = {10.1007/978-3-319-94196-7_51},
  url       = {https://doi.org/10.1007/978-3-319-94196-7_51}
}

@article{rahmanEnergyefficientProjectScheduling2022,
  author  = {Rahman, Humyun Fuad and Chakrabortty, Ripon K. and Elsawah, Sondoss and Ryan, Michael J.},
  title   = {Energy-Efficient Project Scheduling with Supplier Selection in Manufacturing Projects},
  journal = {Expert Systems with Applications},
  year    = {2022},
  volume  = {193},
  pages   = {116446},
  doi     = {10.1016/j.eswa.2021.116446},
  url     = {https://doi.org/10.1016/j.eswa.2021.116446}
}

@article{nguyenGeneticProgrammingProduction2017,
  title = {Genetic Programming for Production Scheduling: A Survey with a Unified Framework},
  shorttitle = {Genetic Programming for Production Scheduling},
  author = {Nguyen, Su and Mei, Yi and Zhang, Mengjie},
  year = 2017,
  month = mar,
  journal = {Complex \& Intelligent Systems},
  volume = {3},
  number = {1},
  pages = {41--66},
  issn = {2198-6053},
  doi = {10.1007/s40747-017-0036-x},
  url = {https://doi.org/10.1007/s40747-017-0036-x},
  urldate = {2024-01-10},
  langid = {english},
}

@article{zhangSurveyGeneticProgramming2023,
  title = {Survey on {{Genetic Programming}} and {{Machine Learning Techniques}} for {{Heuristic Design}} in {{Job Shop Scheduling}}},
  author = {Zhang, Fangfang and Mei, Yi and Nguyen, Su and Zhang, Mengjie},
  year = 2024,
  journal = {IEEE Transactions on Evolutionary Computation},
  volume = {28},
  number = {1},
  pages = {147--167},
  issn = {1941-0026},
  doi = {10.1109/TEVC.2023.3255246},
  url = {https://doi.org/10.1109/TEVC.2023.3255246},
  urldate = {2023-11-26},
}

@inproceedings{lamNumbaLLVMbasedPython2015,
  title = {Numba: A {{LLVM-based Python JIT}} Compiler},
  shorttitle = {Numba},
  booktitle = {Proceedings of the {{Second Workshop}} on the {{LLVM Compiler Infrastructure}} in {{HPC}}},
  author = {Lam, Siu Kwan and Pitrou, Antoine and Seibert, Stanley},
  year = 2015,
  month = nov,
  series = {{{LLVM}} '15},
  pages = {1--6},
  publisher = {Association for Computing Machinery},
  address = {New York, NY, USA},
  doi = {10.1145/2833157.2833162},
  url = {https://doi.org/10.1145/2833157.2833162},
  urldate = {2026-07-30},
  isbn = {978-1-4503-4005-2}
}

@article{hagberg2008,
  author = {Hagberg, Aric A. and Schult, Daniel A. and Swart, Pieter J.},
  title = {Exploring Network Structure, Dynamics, and Function using NetworkX},
  journal = {Python in Science Conference},
  year = {2008},
  doi = {10.25080/TCWV9851},
  url = {https://doi.org/10.25080/TCWV9851},
}

@misc{anthropic_claude_2026,
  author       = {{Anthropic}},
  title        = {Enabling {Claude Code} to Work More Autonomously},
  year         = {2025},
  month        = sep,
  url          = {https://www.anthropic.com/news/enabling-claude-code-to-work-more-autonomously},
  note         = {Accessed: 2026-08-04}
}

@misc{openaiCodex2025,
  author       = {{OpenAI}},
  title        = {{Codex} Is Now Generally Available},
  year         = {2025},
  month        = oct,
  url          = {https://openai.com/index/codex-now-generally-available/},
  note         = {Accessed: 2026-08-04}
}

@article{sarkarVibeCoding2025,
  title   = {Vibe Coding: Programming through Conversation with Artificial Intelligence},
  author  = {Sarkar, Advait and Drosos, Ian},
  year    = {2025},
  journal = {arXiv preprint arXiv:2506.23253},
  doi     = {10.48550/arXiv.2506.23253},
  url     = {https://doi.org/10.48550/arXiv.2506.23253}
}

@article{wilsonGoodEnoughPractices2017,
  title   = {Good Enough Practices in Scientific Computing},
  author  = {Wilson, Greg and Bryan, Jennifer and Cranston, Karen and Kitzes, Justin and Nederbragt, Lex and Teal, Tracy K.},
  year    = {2017},
  journal = {PLOS Computational Biology},
  volume  = {13},
  number  = {6},
  pages   = {e1005510},
  doi     = {10.1371/journal.pcbi.1005510},
  url     = {https://doi.org/10.1371/journal.pcbi.1005510}
}

\end{document}